\documentclass[sigconf]{acmart}

\AtBeginDocument{%
  \providecommand\BibTeX{{%
	\normalfont B\kern-0.5em{\scshape i\kern-0.25em b}\kern-0.8em\TeX}}}

\copyrightyear{2022}
\acmYear{2022}
\setcopyright{acmlicensed}
\acmConference[CHI '22]{CHI Conference on Human Factors in Computing Systems}{April 29-May 5, 2022}{New Orleans, LA, USA}
\acmBooktitle{CHI Conference on Human Factors in Computing Systems (CHI '22), April 29-May 5, 2022, New Orleans, LA, USA}
\acmPrice{15.00}
\acmDOI{10.1145/3491102.3501825}
\acmISBN{978-1-4503-9157-3/22/04}

\usepackage{tabularx}
\begin{document}

\title[Design Guidelines for Text-to-Image Prompting ]{Design Guidelines for Prompt Engineering Text-to-Image Generative Models}

\author{Vivian Liu}
\affiliation{%
  \institution{Columbia University}
  \city{New York}
  \state{New York}
  \country{USA}
}
\email{vivian@cs.columbia.edu}

\author{Lydia B. Chilton}
\affiliation{%
  \institution{Columbia University}
  \city{New York}
  \state{New York}
  \country{USA}
}
\email{chilton@cs.columbia.edu}

\begin{abstract}
Text-to-image generative models are a new and powerful way to generate visual artwork. However, the open-ended nature of text as interaction is double-edged; while users can input anything and have access to an infinite range of generations, they also must engage in brute-force trial and error with the text prompt when the result quality is poor. We conduct a  study exploring what prompt keywords and model hyperparameters can help produce coherent outputs. In particular, we study prompts structured to include subject and style keywords and investigate success and failure modes of these prompts. Our evaluation of 5493 generations over the course of five experiments spans 51 abstract and concrete subjects as well as 51 abstract and figurative styles. From this evaluation, we present design guidelines that can help people produce better outcomes from  text-to-image generative models.
\end{abstract}



\begin{teaserfigure}
\centering
  \includegraphics[width=0.75\linewidth]{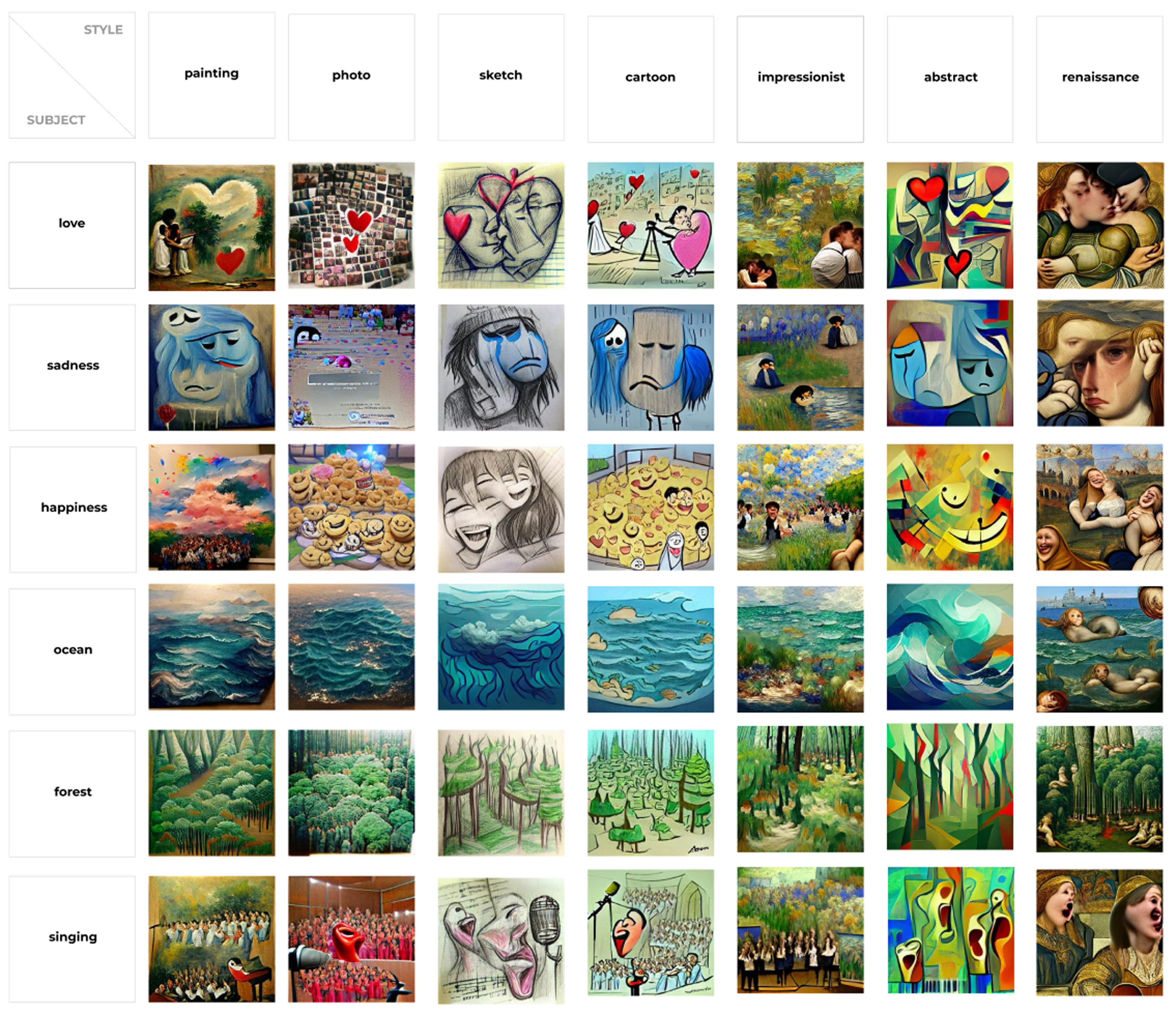}
  \caption{An example grid of text-to-image generations generated from the following prompt template:  "{SUBJECT} in the style of {STYLE}". We analyze over 5000 generations in a series of five experiments involving 51 subjects and 51 styles to study what prompt parameters and hyperparameters can help people produce better outcomes from text-to-image generative models.}
  \Description{An example grid of text-to-image generations generated from the following prompt template:  "{SUBJECT} in the style of {STYLE}" is shown. There are 7 columns of styles and 6 rows of subjects shown. The styles are painting, photo, sketch, cartoon, Impressionist, abstract, and Renaissance. The subjects are love, sadness, happiness, ocean, forest, and singing. Each entry in this subject-style matrix is an text-to-image generation with that subject and style. In this paper, we analyze over 5000 generations in a series of five experiments involving 51 subjects and 51 styles to study what prompt parameters and hyperparameters can help people produce better outcomes from text-to-image generative models.}
  \label{fig:teaser}
\end{teaserfigure}

\maketitle

\section{Introduction}
Recently, advances in computer vision have introduced methods that are remarkable at generating images based upon text prompts \cite{radford2021learning, ramesh2021zeroshot}. For example, OpenAI introduced DALL-E, one such text-to-image model in 2020, and demonstrated that from running a text prompt such as \textit{"a radish dancing in a tutu"}, the model could generate many images matching the prompt.
Based on this progress, artists, programmers, and researchers have come together on communities within Reddit \cite{r/bigsleep} and Twitter \cite{hashtagvqganclip} and developed different models that open source text-to-image generation. Tutorials \cite{tutorialvqganclip,bestiariodelhypogripho_2021} and interactive notebooks \cite{colab} maintained by community members such as @RiverHaveWings, @advadnoun, and @somewheresy on Twitter have made these tools broadly accessible \cite{crowson, adverb, justin}.

Text is free-form and open-ended, so the possibilities for image generation from text prompts are endless. However, this also means that the design process for generating an image can easily become brute-force trial and error. People must search for a new text prompt each time they want to iterate upon their generation, a process that can feel random and unprincipled. In the field of natural language processing, this problem is known as \textit{prompt engineering} \cite{reynolds2021prompt}. Prompt engineering is the formal search for prompts that retrieve desired outcomes from language models, where what is desirable is dependent upon the end task and end user \cite{sanh2021multitask}. There are a number of open questions within prompt engineering to explore for text-to-image models. Some questions relate to hyperparameters: how do variables influencing the length of optimization and random initializations affect model outcomes? Other questions involve the prompt: are there certain classes of words or sentence phrasings that yield better outcomes? These questions are necessary for the HCI community to answer so technical advancements in machine learning such as prompt engineering and multimodal models can be translated into usable interaction paradigms.

To explore the generative possibilities of this system, we systematically approach prompt engineering for a family of prompts that have found traction within practitioners working with text-to image systems: "SUBJECT in the style of STYLE" prompts. In this paper, we address key questions around prompt engineering in a series of five experiments:


\begin{itemize}
    \item \textbf{Experiment 1.} We test different phrasings of the prompt to see how modulating the language of the prompt with different orderings, function words, and filler words affects generation quality. 
    \item \textbf{Experiment 2.} We test different random initializations to find an optimal range of generations to produce for each prompt, accounting for the probabilistic behavior of text-to-image frameworks.
    \item \textbf{Experiment 3.} We vary and study the number of iterations to find an optimal range for the length of optimization.
    \item \textbf{Experiment 4.} We explore styles as a parameter of the prompt to understand the breadth of styles the system can reproduce. Specifically, we explore 51 styles spanning different time periods (modern vs. premodern vs. digital), schools of culture (Western vs. non-Western), and levels of abstraction (abstract vs. figurative). Additionally, we look for biases across these different partitions of styles.
    \item \textbf{Experiment 5.} We explore subjects as a parameter of the prompt to understand how subject and styles interact with each other. We tested 51 subjects across 31 styles to explore whether the system is better at producing abstract subjects or concrete subjects given an abstract or a figurative style.
\end{itemize}

In Experiments 4 and 5, we provide qualitative analysis of observed success and failure modes one might encounter while working with text-to-image generation. We conclude with design guidelines to help end users prompt text-to-image models for observed success modes and steer away from observed failure modes.

\section{Related Work}

\subsection{Generative Methods as Creativity Support Tools}

Artist and programmer communities have consistently shown interest in the potential of generative AI as an art medium. Communities conversing about artistic AI have developed for a long time around networks such as DeepDream \cite{mordvintsev_tyka_olah_2015}, neural style transfer networks \cite{gatys2015neural}, and generative adversarial networks \cite{goodfellow2014generative,karras2020analyzing, brock2019large}. Likewise, HCI researchers have sought to understand how generative AI can become a creativity support tool for artists. In recent years, systems embedding generative models have been successfully applied to domains such as image generation, poetry, and music \cite{ghosh2019interactive, 10.1145/3411764.3445093,10.1145/3290605.3300526, huang2020ai}.

Often, researchers try to leverage the large space of design solutions that generative methods provide to assist users during ideation and iteration \cite{10.1145/3313831.3376739}. However to do so, researchers have to understand how users can explore these design solution spaces efficiently and effectively. This is an open question that has been investigated through a number of research approaches. For example, Matejka et al. \cite{Matejka2018} introduced Dreams Lens, a system implementing design galleries and interactive data visualization to visualize diverse design solutions from a 3D modeling generative program. Yang et al. \cite{liu2019-lsc} proposed latent space cartography, which used dimensionality reduction to explore the latent design space of generative AI models. Shimizu et al. \cite{shimizu} proposed Design Adjectives, a system that helped users parameterize the design space by first giving examples of what attributes they liked and disliked for the design of fonts, materials, and motion graphics .

One drawback of many of these AI-based approaches is that while they can create an inexhaustible number of generations,  they lack meaningful and interpretable controls for users. This problem has given rise to an area of research at the intersection of HCI and AI focused on semantically meaningful exploration. One of the earliest works in this direction was a seminal creativity support system called AttriBit, which allowed users to assemble 3D models given data-driven suggestions. These suggestions supported semantic goals users crafted for their creations (i.e. creating  a “cute” or “dangerous” 3D animal) \cite{10.1145/2501988.2502008}. More recent work leveraging deep learning has also tried to produce semantically meaningful editing operations. For example, Louie et al. \cite{10.1145/3313831.3376739} introduced CoCoCo, an AI music creation system which lets users move sliders to tune their generated music to be "happier" or "sadder". Geppeto was another analogous mixed-initiative, co-creative system that generated robot animation according to mood-related semantic goals \cite{10.1145/3313831.3376739, Desai:2019:GES:3290605.3300599}. Systems that can directly support user goals in a semantically meaningful way are both more interpretable and more usable.

\subsection {Text-to-Image Generation}
This interest in involving semantics and natural language as a form of interaction with generative models has recently found success in machine learning. Recent work within representation learning has focused on learning text and image understanding together by coupling the two modalities through a contrastive objective during optimization. In 2021, Radford et al. \cite{radford2021learning} from OpenAI introduced CLIP, a method for learning multimodal image representations. CLIP was trained on an Internet scale size dataset of 400 million image and text pairs to learn a multimodal embedding space that incorporated both text and image understanding. CLIP demonstrated that the model was able to learn "visual concepts...enabling zero-shot transfer of the model" on various tasks such as OCR, geo-localization, and others. CLIP was used in DALL-E \cite{ramesh2021zeroshot}, one of the state of the art models for text-to-image generation. DALL-E learned a transformer that autoregressively predicted text and image tokens together in one sequence. The authors of DALL-E demonstrated how the model could handle image operations, perform style transfer, and produce novel combinations of elements.
An outcropping of text-to-image architectures achieving similar functions followed: DMGAN \cite{zhu2019dmgan}, VQGAN+CLIP \cite{nerdyroden},BigSleep (BIGGAN+CLIP) \cite{murdock1}, DeepDaze (SIREN+CLIP) \cite{murdock2}, CLIP-guided diffusion  \cite{crowsondiffusion}. Many models were open sourced and advanced within the creative technologist community.

\subsection{Prompt Engineering}

Researchers and practitioners alike now tackle the open problem of prompt engineering for large pretrained models. Most work in prompt engineering has concentrated within the text generation problem from natural language processing. The term prompt engineering originally came from a popular post online about GPT3 (a large language model) and its capabilities for writing creative fiction. The author, Gwern Branwen \cite{gwern_2020}, suggested that prompt engineering models could become a new paradigm for interaction; users need only figure out how to prompt a model to elicit the specific knowledge and abstractions necessary for completing downstream tasks. Follow-up work from practitioners have disseminated prompt engineering methods and tricks such as prefix-tuning and using few-shot examples \cite{cantino_2021}.

This paradigm was formalized by Liu et al. \cite{liu2021pretrain}, who referred to this emerging paradigm as "pretrain, predict, and prompt". They further enumerated a schema for prompt templates categorizing prompts based on prompt shape (cloze and prefix prompts), answer engineering (answered and filled prompts), and task-specific prompts (i.e. prompt templates tailored for tasks like summarization or translation). Additionally, they expanded on alternative approaches to prompt engineering such as automated template learning and multi-prompt engineering.

While momentum has started to build in prompt engineering for \textit{text} generation purposes, less work has been done to rigorously examine how users can prompt generative frameworks with natural language for \textit{visual} generation purposes, which is the focus in this paper. To our knowledge, one of the few works close to ours is by Ge and Parikh et al. \cite{ge2021visual}, who utilize BigSleep (BigGAN+CLIP) and DeepDaze (SIREN+CLIP) for visual conceptual blending. Their approach used BERT \cite{devlin2019bert} to generate prompts and help users make visual blends, using shape keywords to prime the generation.

So far, progress on prompt engineering for visual tasks and end user usability has been made informally and in an ad hoc fashion. Creative technologists have discussed tricks and keywords that help tune models towards their aesthetic goals. For example, Aran Komatsuzaki, a prominent artist and research programmer noted that using 'unreal engine' as a prompt helped them add a hyperrealistic, 3D render quality to their image generation \cite{komatsuzaki_2021}. This tweet and many others along the same vein established a growing trend within the artistic community to structure prompts with the template \textit{"X in the style of Y"} , where Y would be an artist or art movement that CLIP would ideally have knowledge of. In the experiments in this paper, we evaluate this family of prompts to systematically conduct prompt engineering.

\subsection{Probing through Prompt Engineering}
Literature has shown that evaluating a constrained set of keywords and prompts can help better explain and interpret learned models. For example, influential work by Caliskan et al. \cite{2017} used sets of words to quantifiably demonstrate bias within word embeddings. Specifically, they studied the GloVE word embedding and showed that small sets of gendered words significantly correlated with attribute words, identifying associations such as female-gendered words with family-oriented words and male-gendered words with career-oriented words. These experiments helped formulate a global understanding of a computational model and the biases embedded within them.

A significant amount of work has also gone into probing and interpreting what large pretrained models learn and utilize at inference time. Work on BERT such as “A Primer on BERTology” \cite{rogers2020primer} and “BERT Relearns the Classical NLP Pipeline” \cite{tenney2019bert} have probed what BERT learns across its layers and what world knowledge it holds within. For example, \cite{rogers2020primer} states that BERT “struggles with abstract attributes of objects as well as visual and perceptual properties are assumed rather than mentioned.”

It is important to apply this direction of research to multimodal models such as CLIP (which is a key component within VQGAN+CLIP and multiple other text-to-image generation frameworks) and to understand what CLIP holds within its knowledge distribution. Understanding the local behavior and global knowledge distributions of AI models can help users develop better mental models of them as agents \cite{gero2020}. Using prompts to generate image evidence of AI knowledge is also a way of reducing uncertainty with AI \cite{10.1145/3313831.3376301}. Prompt engineering thus is both a human-computer interaction paradigm to support as well as a valuable method of probing deep models.

\section{Experiment 1. Prompt Permutations}
In language, there are many ways to say the same thing in different words. We wanted to understand the effect this in the context of text-to-image generation. Would users need to try many different permutations of the same prompt to get a sense of what a prompt would return, or would just one suffice? Additionally, would there be certain permutations of the prompt keywords that would lead to better generations and be the best way to word a prompt? For example, would prompting the model with "a woman in a Futurist style" lead to a significantly different generation than "a woman painted in a Futurist style", "woman with a Futurist style", or "a woman. Futurism style"? In this experiment, we wanted to rigorously examine the following question: \textbf{do different rephrasings of a prompt using the same keywords yield significantly different generations?}

Our original hypothesis about this question was that there would be no prompt permutation that would do significantly better or worse than the rest, because none of the rephrasings seemed to have significantly more meaning than the next.

\subsection{Methodology}
To study different permutations of prompts, we first had to generate a large number of images. To do this, we used the checkpoint and configuration of VQGAN+CLIP pretrained on Imagenet with the 16384 codebook size \cite{nerdyroden}. Each image was generated to be 256x256 pixels and iterated on for 300 steps on a local NVIDIA GeForce RTX 3080 GPU.

Each image was generated according to a prompt involving a subject and style. We chose the following subjects: \textit{love, hate, happiness, sadness, man, woman, tree, river, dog, cat, ocean, and forest}.  These subjects were chosen for their universality across media and across cultures. These subjects additionally were balanced for how abstract or concrete they were as a concept as well as for positive and negative sentiment. We decided on whether a subject fell into the abstract or concrete category based upon ratings taken from a dataset of concreteness values \cite{article}. Our set of abstract subjects averaged 2.12 on a scale from one to five (one being most abstract), and our set of concrete subjects averaged 4.80.

Similarly, we chose 12 styles spanning different time periods, cultural traditions, and aesthetics: \textit{Cubist, Islamic geometric art, Surrealism, action painting, Ukiyo-e, ancient Egyptian art, High Renaissance, Impressionism, cyberpunk, unreal engine, Disney, VSCO}. These styles likewise varied in whether they represented the world in an abstract or figurative manner. Specifically, we chose four abstract styles, four figurative styles, and four aesthetics related to the digital age. We balanced for time periods (with 6 styles predating the 20\textsuperscript{th} century, and 6 styles from the 20\textsuperscript{th} and 21\textsuperscript{st} century).

We used these 12x12 subject and style combinations to study the effect of prompt permutations: how different rephrasings of the same keywords affect the image generation. For each of these combinations, we tested 9 permutations derived from the the CLIP code repository and discussion within the online community, generating 1296 images in total. The nine permutations are as follows, and the specific rationale for each permutation is listed in the Appendix:

 \begin{itemize}
   \item \textbf{A {MEDIUM} of {SUBJECT} in the {STYLE} style }  --- Example: a painting of love in the abstract style
 \item \textbf{A {STYLE} {MEDIUM} of a {SUBJECT}} --- Example: an abstract painting of love
\item \textbf{{SUBJECT} STYLE} --- Example: love abstract art
\item \textbf{{SUBJECT}. STYLE} --- Example: love.abstract art
\item \textbf{{SUBJECT} in the style of {STYLE}} --- Example: love in the style of abstract art
\item \textbf{{SUBJECT} in the {STYLE} style} --- Example: love painted in the abstract art style
\item \textbf{{SUBJECT} {VERB} in the {STYLE}} style --- Example: love painted in the abstract style
\item \textbf{{SUBJECT} made/done/verb in the {STYLE} art style} --- Example: love done in the abstract art style
\item \textbf{{SUBJECT} with a {STYLE} style.} --- Example: love with an abstract art style

 \end{itemize}

\subsection{Annotation Methodology}
We had each combination of subject and style rated by two people who had backgrounds in media arts and art practice respectively. The 144 subject and style combinations were presented in 3x3 grids, where the prompt permutations were randomly arranged to prevent any effect from ordering. One combination was taken out owing to inappropriate content.

Annotators were asked to note which images in the grid were either significantly better generations or significantly worse generations. We explained that they did not have to judge whether or not the generation represented the subject or the style; they just had to report whether there were generations that were significantly different from the rest. For example, if there was a different element that emerged in one generation or a shift in color palette compared to the rest---these differences constituted outliers. All annotators were compensated \$20/hour for however long it took them to complete the task. This rate of compensation was the same for the rest of our experiments.

\begin{figure*}[t]

	\includegraphics[width=\linewidth]{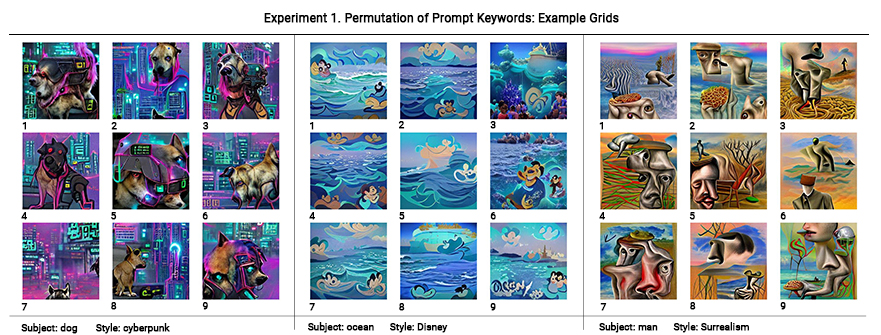}
	\caption{For Experiment 1, annotators judged 3x3 grids where generations from different prompt permutations were arranged randomly. Annotators evaluated 143 grids of generations for significantly better generations as well significantly worse generations (outliers in generation quality). We found no significant difference between the quality of the images that these nine prompt permutations generated, and therefore no significant difference between different prompt permutations.}
	\Description{Three by three grids are shown, illustrating what annotators saw in the prompt permutations task. These grids show generations that vary slightly in terms of composition but which share great similarity in terms of color palette and overall aesthetic. For Experiment 1, annotators judged 3x3 grids where generations from different prompt permutations were arranged randomly. Annotators evaluated 143 grids of generations for significantly better generations as well significantly worse generations (outliers in generation quality). We found no significant difference between the quality of the images that these nine prompt permutations generated, and therefore no significant difference between different prompt permutations.}
	\label{fig:permutations}
\end{figure*}

\subsection{Results}

From the annotations we collected, we binned the generations based upon whether they were annotated as the same as the rest of the group or marked as an outlier. Outliers were generations that were  either "significantly better" or "significantly worse". After aggregating across these two categories, we checked agreement between our two annotators. We observed high agreement, at 71.3\% across 1296 generations. We then calculated interrater reliability, where we observed a Cohen's kappa of 0.0013. This value is low, but we believe this comes from the subjective nature of the task. We can see this in the example grids of \autoref{fig:permutations}, which are composed of slightly varying generations. While we provided examples of what might constitute a significantly different generation and modeled the task for annotators as best we could, picking outliers is still inherently subjective and this subjectivity could have influenced the factor calculated in Cohen's kappa that models chance. Therefore, even though our Cohen's kappa value was low, we proceed based on the high agreement value of 71.3\% across 1296 generations.
	
We assembled a contingency table based upon the following categories of annotation possibilities, same-same, same-outlier, outlier-same, and outlier-outlier. We performed a Chi-square test based upon this contingency table. We found that with a Chi-squared test statistic of 0.354 and a p-value of 0.55, the number of prompt permutations judged as outliers was insignificant when compared to the number of prompt permutations deemed not outliers.



Hence, we concluded that there was no significant difference between the nine prompt permutations that we tried. We synthesized the following guideline from this experiment: \textbf{When picking the prompt, focus on subject and style keywords instead of connecting words.} The connecting words (i.e. function words,  punctuation, and words for ordering) did not contribute statistically meaningful differences in generation quality. Hence, we moved forward in the following experiments testing only one prompt permutation per subject and style combination rather than multiple rephrasings for the same combination.

\section{Experiment 2. Random Seeds}

A common parameter in generative models is seeds. Generative models are stochastic and highly dependent upon their initializations, which means that it is often hard to reproduce results. To mitigate this, people often use seeds to have reproducible results and behavior. We noticed that using different seeds with VQGAN+CLIP resulted in generations that would differ in composition. We wanted to understand: \textbf{do different seeds using the same prompt yield significantly different generations?} The motivation behind this question was to understand whether or not users would need to try multiple seeds before moving onto new combinations of keywords.

Our hypothesis was that no seed would do significantly better or worse than the rest, because changing seeds and altering the random initialization of the model should not produce any consistent or significant signal.

\subsection{Generation Methodology}

To study the effect of seeds, we generated 1296 images from 12 subjects, 12 styles, and 9 seeds. Because neither subject nor style were the main focus of this experiment, we chose to use the same set of subjects justified in the previous section. Likewise, we chose to use the same set of styles.
What we did vary was the seed chosen. We generated images using 10 randomly generated seeds 796, 324, 697, 11, 184, 982, 724, 962, and 805 and the prompt ``{SUBJECT} in the style of {STYLE}''.
\subsection{Annotation Methodology}
Two annotators were shown 1296 generations in 3x3 grids where the seeds were arranged randomly. We had the nine generations varied by seed for each combination of subject and style rated by two people. Annotators were again asked to note which images in the grid were significantly better generations and significantly worse generations from the rest of the group, if any.

\begin{figure*}
	\includegraphics[width=\linewidth]{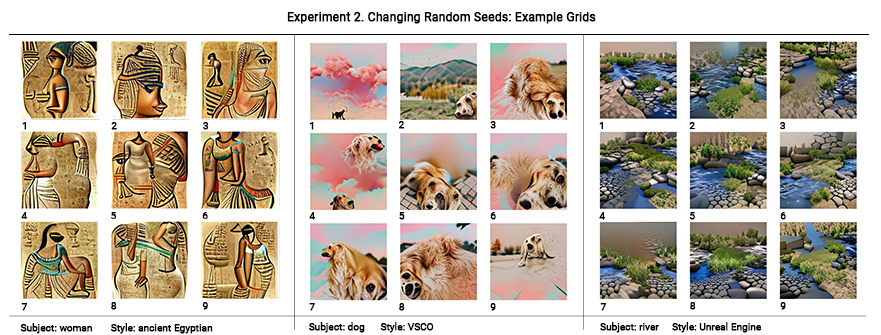}
	\caption{For Experiment 2, annotators judged 3x3 grids such as the ones above where generations utilizing different seeds were arranged in random order. These 143 grids were judged for significantly better generations as well as significantly worse generations. We found that the number of generations judged as outliers in generation quality was significant, meaning that the choice of initializing seed can significantly vary the quality of the generation.}
	\Description{Three three by three grids are shown illustration example grids annotators saw for the experiment evaluating random seed initializations. These generations varied slightly in composition but shared many aesthetic characteristics. For Experiment 2, annotators judged 3x3 grids such as the ones above where generations utilizing different seeds were arranged in random order. These 143 grids were judged for significantly better generations as well as significantly worse generations. We found that the number of generations judged as outliers in generation quality was significant, meaning that the choice of initializing seed can significantly vary the quality of the generation.}
	\label{fig:seed}
\end{figure*}

\subsection{Results}
We again used a Fisher's exact test to evaluate how many generations were judged as approximately the same versus how many were judged outliers. We found that with a p-value of <0.01, the number of generations judged as outliers was significant when compared to the number of generations deemed not outliers. Our annotators shared an inter-rater reliability of 0.13, which indicates slight agreement, which we again justify as valid given the highly subjective nature of the task (picking 'better' or 'worse' images).
This result was surprising to us, because it demonstrated that even outside of the prompt, there are stochastic components of the generation that can significantly vary the quality of the generation. We conclude from this experiment that it is prudent to try multiple seeds during prompt engineering. A design guideline that follows is to \textbf{ generate between 3 to 9 different seeds to get a representative idea of what a prompt can return.}.

\section{Experiment 3. Length of Optimization}

A free parameter during each run of text-to-image models is the length of optimization: the number of iterations the networks are run for. Typically, we can expect that the more iterations, the lower and more stable the loss, and ideally the better the image. We wanted to investigate on average how many iterations are needed to get a decent result. We also wanted to see if runs with lower iterations could produce images with just as good generation quality as runs with higher iterations; a lower number of iterations means faster results, and for future systems involving text-to-image generation, we would want to know an average number of iterations needed to arrive at a favorable result. Our specific research question was: \textbf{does the length of optimization correlate with better evaluated generations?}


\subsection{Generation Methodology}

To investigate this, we tested 6 subjects (\textit{happiness, sadness, man, cat, ocean, and forest}) across 12 styles, with a constant seed and one variety of prompt permutation. We ran the generations for 1000 iterations, and had users evaluate the generations every 100 iterations. We chose 1000 iterations as the maximum because we wanted to try a number of short to moderate wait times and 1000 is a suggested default.

\subsection{Annotation Methodology}

\begin{figure*}[t]
	\centering
	\includegraphics[width=\linewidth]{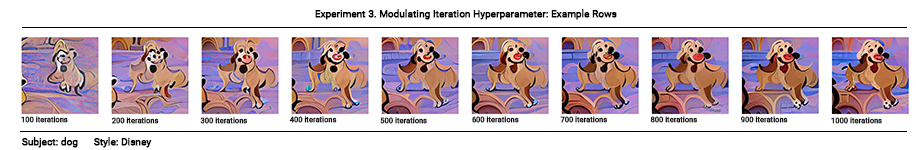}
	\caption{ For Experiment 3, annotators were shown rows such as the row above. These images represented different iteration steps of the optimization process. Annotators chose the iteration step that they most preferred from these sets of 10.}
	\Description{A row of generations shows the generation at different stages of iteration. Each entry progresses 100 iterations starting from the 100th iteration to the 1000th. The subject is a dog, and the style is Disney. Earlier iterations have a softer and noisier quality, while later iterations have stronger contrast.}
	\label{fig:iterations}
\end{figure*}

We had annotators annotate rows of generations saved at different steps of the iteration. These were specifically steps that were multiples of 100 up to 1000. The 0\textsuperscript{th} iteration was not shown because the generation always began from random noise. Annotators annotated 72 rows for which generation they most preferred from the set of 10.

\subsection{Results}

\begin{figure}[t]
	\centering
	\includegraphics[width=0.8  \linewidth]{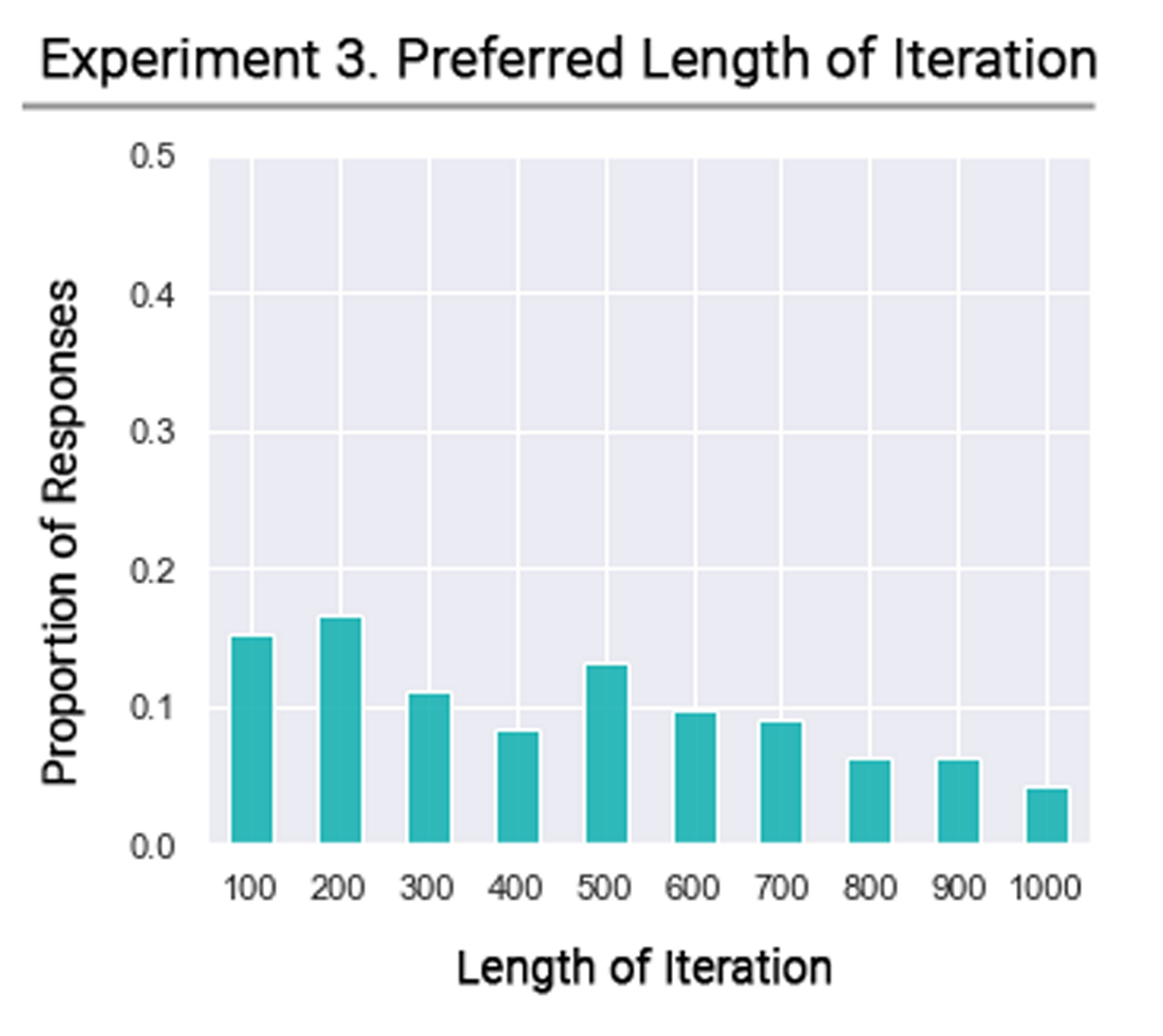}
	\caption{Above is a plot for Experiment 3 of the frequency of times annotators chose an iteration step as their favorite iteration. Lower values of iterations (100-500) tended to be preferred, and this difference over higher values of iteration was significant.}
	\label{fig:iterationsfigurestats}
\end{figure}

We found that the differences between the chosen iteration steps were significant upon performing a chi-squared test (p-value=0.01). We include the observed frequencies for the preferred iteration steps in \autoref{fig:iterationsfigurestats}, where we can see that 200, 100, and 500 iterations being chosen as the most preferred. We report a Cohen's kappa score of 0.33, which represents fair agreement, which we think is valid considering the highly subjective nature of picking a preferred iteration step (but more conducive to agreement than picking intuitive favorites in Experiment 2, which evaluated seeds).

This demonstrates that a higher number of iterations did not necessarily correlate with a more desirable generation, as one might have been expected considering that more iterations optimize the image and text representations towards one another. This is a non-intuitive result meaning that the current multimodal methods are not necessarily optimizing for generations that we \textit{prefer}. Possible explanations could include the fact that lower iterations tended to have a softer quality compared to higher iterations, where the differences and contrast seemed to be more exaggerated in higher iterations.

We conclude from this experiment \textbf{when generating with fast iteration in mind, using shorter lengths of optimization between 100 and 500 iteration is sufficient}. However, as can be seen in \autoref{fig:iterations}, at the lower end of this range (ie. 100 iterations), the subject is not guaranteed to have manifested in the generation yet. One reason people preferred 100 or 200 iteration generations is because in certain styles such as abstract ones, the subject does not need to manifest as saliently. Therefore, we suggest 300 iterations as a good default, which is also what we use throughout the rest of the experiments.

\section{Experiment 4. Testing a Breadth of Styles}

We understood that style was a keyword we could use to suggest an aesthetic within a generation. However given that there are an abundant amount of styles, we wanted to see if the model would perform equally well across a breadth of styles. Could VQGAN+CLIP, in all its pretraining, handle \textit{any} style? In addition, we also wanted to understand if the framework would perform differently for different classes of styles and if the framework was biased towards certain traditions of styles.

To rigorously investigate this, we looked at three ``partitions'' of styles: abstract versus figurative, Western versus non-Western, and styles partitioned by time period (digital, modern, and premodern).

We had the following hypotheses for each:

\begin{itemize}
	\item For abstract versus figurative styles, we assumed abstract styles would perform better because we thought they would be more tolerant to the deconstructed, global incoherence endemic to many generations.
	\item For Western versus non-Western styles, we assumed that the model would perform better on Western art, since many of the computer science datasets relevant to art focus on Western schools of art (i.e. WikiArt and MetFaces) \cite{www.wikiart.org, Tero2020}.
	\item For styles partitioned by digital, modern, and premodern time periods, we thought digital styles would do better, as the model we used was trained on images and text from the Internet.
\end{itemize}

\subsection{Generation Methodology}
To comprehensively investigate the breadth of stylistic knowledge the framework had within reach and to see if we could use style as a keyword to structure prompts, we tested a large number of styles.

A style, from the perspective of art history, represents a distinctive way in which visual arts can be grouped. To operate on this definition methodologically, we pulled styles from existing knowledge bases of art history and aesthetics online. We looked in particular at The Metropolitan Museum of Art Heilbrunn Timeline of Art History, the Aesthetics Wikia, the schema by which the WikiArt dataset was organized, and relevant Wikipedia articles to produce the set of 51 styles enumerated in \autoref{styleslisted} \cite{metmuseum.org, fandom}. These styles were chosen to balance certain factors that influence style such as time periods, culture, and whether they were abstract or figurative in the way the style represented the real world.

\begin{table*}

\footnotesize
\begin{tabularx}{\linewidth}{c | c | c | c | c | c | c }
\toprule
Medium & West Fig. Pre. & West Fig. Mod. & Non-West Abs. Mod. & Non-West Abs. Mod. & West Abs. Mod. & Internet | Aesthetics \\
\midrule
 painting & Baroque & Pop Art & Ukiyo-e& Mola art & action painting & fractal\\
 photo & High Renaissance & Surrealism & Chinese ink wash painting& Geometric Islamic art & Op art &VSCO  \\  
 sketch & Impressionism & documentary photography &Kerala mural & Mexican Otomi & Bauhaus & unreal engine 	\\
 cartoon & Medieval &Art deco & Mayan art & Andean textile & Cubism& ASCII art \\
 icon & Pointillism & Hippie movement &African masks& Aboriginal art & Dadaism & Disney  \\
 vector &Neoclassicism  & photorealism & ancient Egyptian art &  & Futurism & Studio Ghibli
 \\
  graffiti & & & thangka &  & & glitch \\ 3D render & & & & & & Cottagecore \\ & & & &  & & Dark academia \\ & & & &  & & Cyberpunk \\ & & & &  & & Pixar \\ & & & &  & & Pokemon \\
\bottomrule

\end{tabularx}

\caption{In Experiment 4, we generated generations from 51 styles listed in the table above. Eight styles were general mediums of visual art. Twelve styles were Internet aesthetics. The remaining 33 were styles that were balanced for representation on both sides of the following partitions: abstract-figurative, Western and non-Western,  and lastly premodern, modern, and digital.}
\label{styleslisted}
\end{table*}

\subsection{Annotation Methodology}
Two annotators with backgrounds in media arts and art practice each received a set of subject and style combinations in random order. They additionally received links to Google Images for each style, in case they needed visual references for styles. They annotated each generated combination (which was just a single image) as per the following rubric:

\begin{itemize}
	\item 1: Extremely poor representation of the style, no motifs were present
	\item 2: Bad representation of the style, few motifs were present
	\item 3: Average representation of the style, some motifs were present
	\item 4: Good representation of the style, high number of motifs were present
	\item 5: Excellent representation of the style, very high number of motifs were present

\end{itemize}

Each annotator was instructed to judge how well the style was represented, irrespective of how well the subject turned out in the generation.

\subsection{Results}

\begin{figure*}[t]
	\centering
	\includegraphics[width=\linewidth]{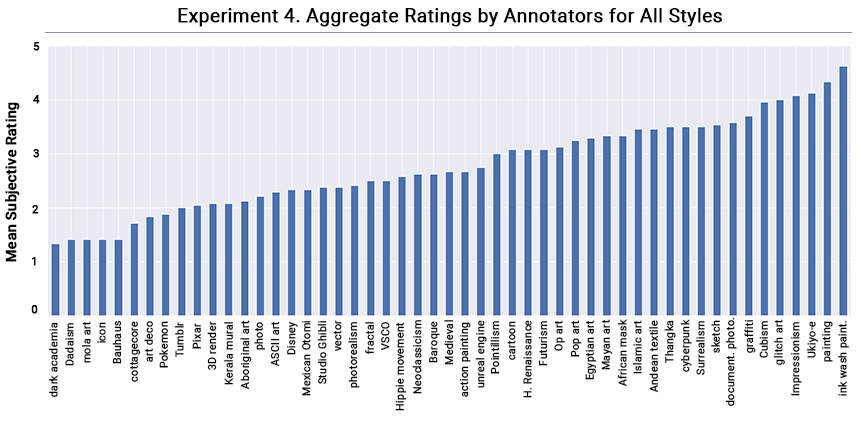}
	\caption{For Experiment 4, 51 styles were tested across 12 subjects. This plot aggregated the ratings across all subjects in a style and ranked the styles from low to high for mean subjective rating.}
	\Description{A bar graph shows how 51 styles performed with the lowest ranking styles on the left and the highest ranking styles on the right. The X axis is styles studied, and the Y axis is the mean subjective rating of the annotators.}
	\label{fig:allstyles}
\end{figure*}

\begin{figure*}
	\centering
	\includegraphics[width=\linewidth]{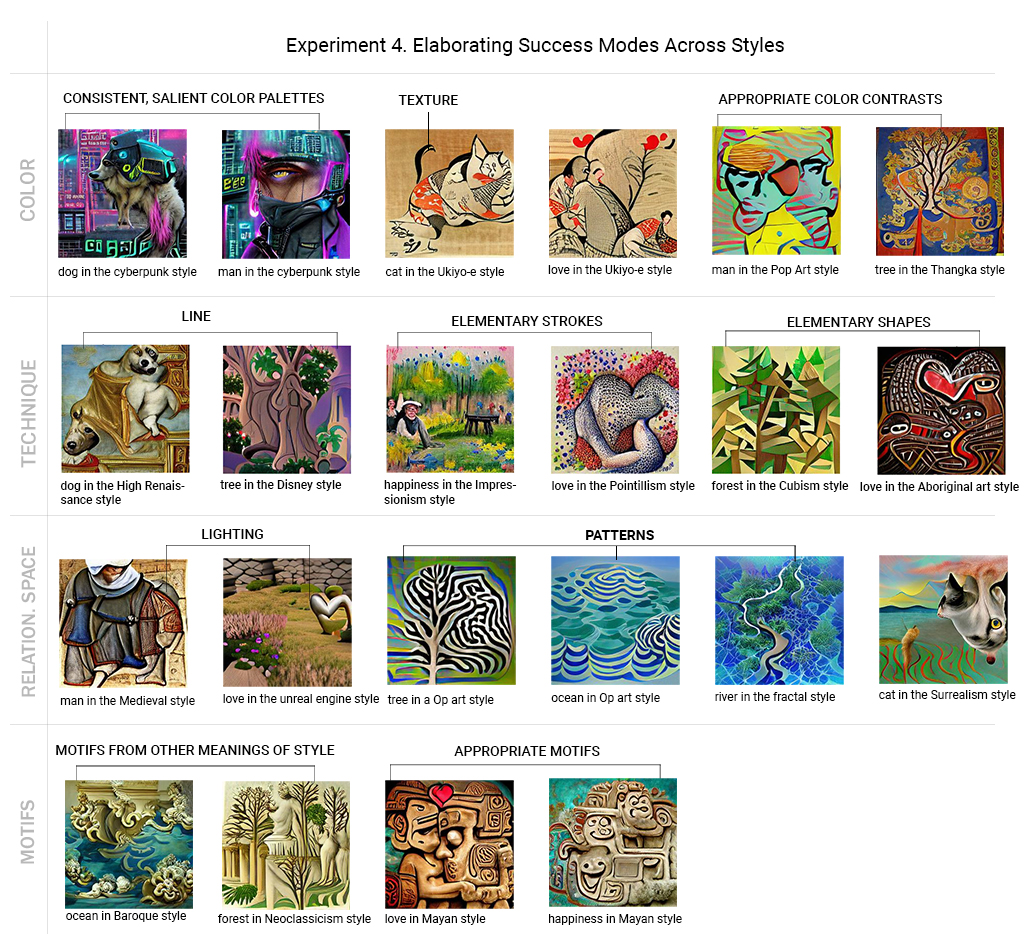}
	\caption{In Experiment 4, we qualify different modes of success seen across the 51 styles we generated for. These modes were \textit{color, technique, relationships in space,} and \textit{motifs}. Generations were able to express \textit{color} well in terms of stereotypical color palettes, textures, and contrasts. The basics of \textit{technique} were also captured in the variety of lines, elementary strokes, and shapes expressed. Generations also demonstrated proficiency in setting lighting and forming patterns, establishing good \textit{relationships in space}. \textit{Motifs} from styles were also readily accessible; for example, we can see ornate swirls in Baroque generations and the dimensional features relevant to Mayan relief sculptures in the last row as motifs of their styles.}
	\Description{In this image we show in four rows examples of the different success modes we delineate within the results of Experiment 4. In the first row for colors, we look at two example generations from the styles of cyberpunk, Ukiyo-e, and thangka. In technique, we look at the styles of  High Renaissance, Pointillism, Impressionism, Cubism, and the Aboriginal art style. In the relationships in space row, we look at successful examples of lighting and patterns achieved by prompts like "ocean in Op art style" and "love in the unreal engine style". In the last row of motifs, we look at generations such as "love in Mayan style", which has the relief sculpture motifs of Mayan art and "ocean in Baroque style", which has the ornate swirls of Baroque work.}
	\label{fig:successcolor}
\end{figure*}

\autoref{fig:allstyles} shows average ratings for all 51 styles and illustrated that the model performed better on some styles than others. We first elaborate on the success modes and failure modes across all the styles as a whole before approaching the partition experiments in depth. Refer to \autoref{fig:successcolor} for the visual depictions of the success modes we observed.

\subsection{Success and Failure Modes for Styles}

\subsubsection{Success Mode. Salient color palettes and relevant textures.}

The first recurring theme across successful styles was the presence of salient color schemes. This was apparent in some of the most positively judged styles such as \textit{Ukiyo-e, glitch art, cyberpunk,} and \textit{thangka}. These generations, pictured in \autoref{fig:successcolor},  demonstrate that text-to-image models can match styles to some of their signature color palettes without explicitly involving color details in the prompt. \textit{Cyberpunk} consistently returned a global aesthetic dominated by halogen colors like cyan and magenta, and \textit{Glitch art} always pulled together colors reminiscent of TV static. Likewise, in generations such as \textit{"tree in a thangka style"} or \textit{"man in the pop art style"}, we see different but correct understandings of the way primary colors can be saturated, contrasted, and complemented.

Texture was another element that came across in many styles. The most successful style seen from the annotations was \textit{ink wash painting}. All generations of ink wash painting were done in wide swathes of ink that captured the watercolor quality of ink on paper. In many premodern styles such as Ukiyo-e, ancient Egyptian, Medieval, the textures of aged paper and papyrus backgrounding the image as hints of canvas also helped express the style.

\subsubsection{Success Mode. Technique}

Another theme across successful styles was the emulation of correct technique. Many generations exhibited choices of \textit{line, texture,} and \textit{elementary brush strokes} congruent with their style.

Across generations of the same style, the model showed the ability to use correct and consistent choices of lines. For styles such as \textit{sketch}, the model produced thin lines suggestive of pencil, while for styles such as \textit{Disney} or \textit{Pokemon}, the model consistently produced thick black outlines characteristic of cartoons. These lines hardly appear at all in styles such as Impressionism, which were composed instead of a patchwork of small and short strokes reminiscent of the style's broken-color technique.

In Impressionism as well as other styles like \textit{Pointillism} and \textit{Cubism}, the model showed its ability to find the right style primitives. \textit{Pointillism} was composed of dots and points and \textit{Cubism} of deconstructed shapes. In certain styles, however, such as \textit{aboriginal art} which often uses dots as their primitives, the model was only able to generate textured patterns suggestive of the dot primitives.

\subsubsection{Success Mode. Depicting Space}
 
The model was generally able to capture the right perspective, which we refer to as whether the image was done in two dimensions or three and how light and shadow were represented in the image. For example, in the Medieval style, light and shadow tended to come across flatly, while on the other extreme in styles such as unreal engine, the 3D scene lighting was very apparent in the pronounced light glares and raycast quality of reflections off elements like metal or hair.

In looking at how the generations depicted space we also assessed composition. We found that styles where patterns and deconstructed gestalts were common were rated favorably. For example, the alternating patterns of swirling and swelling black and white strips canonical to Op art were present in all Op art generations. Other examples of this success mode include the multiplicity of shapes in Cubism, recursive details in fractals, and concentric variegation in aboriginal art. Figurative styles with a high tolerance for deconstructed objects such as Surrealism also performed well in the ranked annotation. This success mode could have potentially been influenced by the convolutional components of VQGAN+CLIP’s architecture. Convolutional representations are inherently focused on local neighborhoods, and this could have been latently optimized for in the patterns and repetition we observed. 
\subsubsection{Success Mode. Motifs of the style}
Certain styles placed motifs, or distinctive details, within the generation that could immediately evoke the style. This was especially true in styles such as the Baroque style, for which the model constantly incorporated lavish details such as ornate swirls and heavy moldings, or the Neoclassicism style, for which the model generated Grecian pillars and drapery within the shapes and contours of the image. It is interesting to note that while these motifs are relevant to the styles, they borrow from different facets of the meanings for Baroque and Neoclassicism that do not necessarily represent the visual arts version of the style that we had intended. We explore the ways the model misunderstood styles in the next section on failure modes.

\subsection{Failure Modes}

\begin{figure*}
	\centering
	\includegraphics[width=\linewidth]{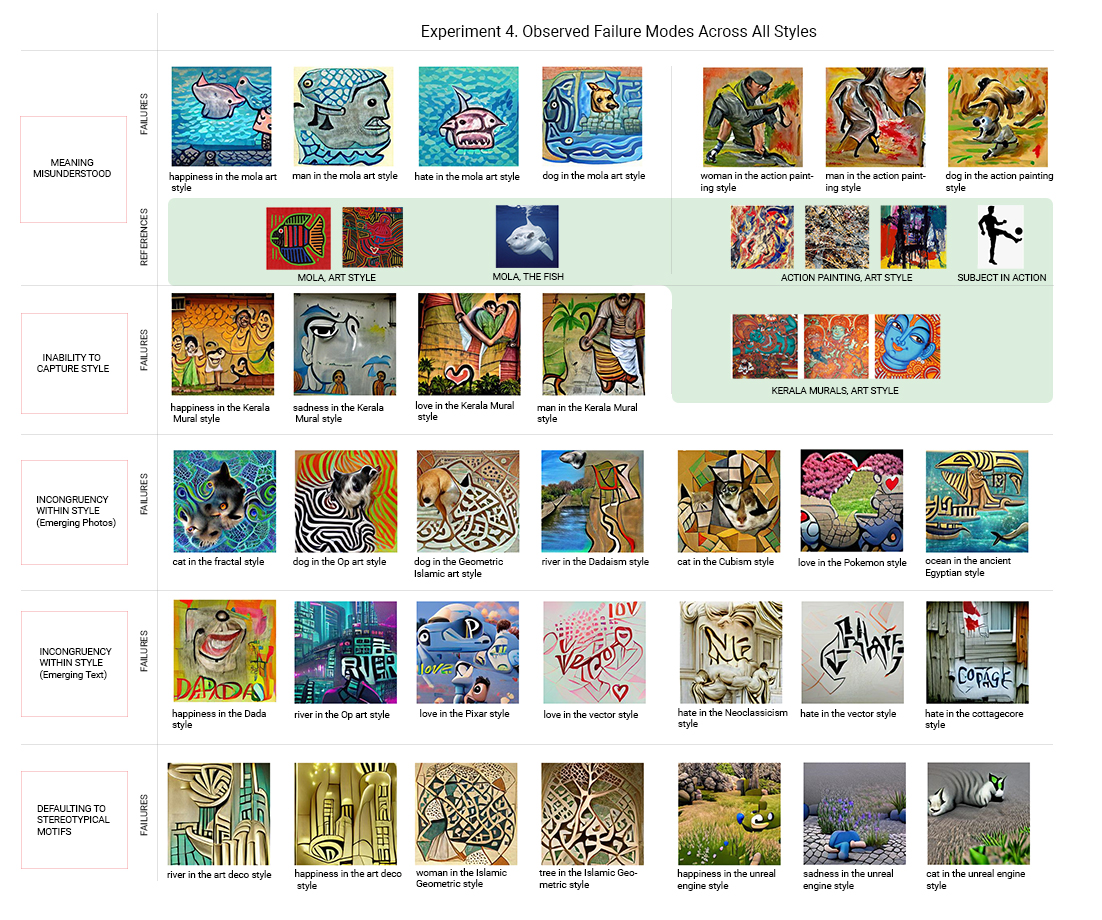}
	\caption{For Experiment 4, we illustrate the failure modes we observed across styles: misunderstandings owing to the multiple interpretations of the text prompt, an inability to correctly capture the style, style incongruencies, and defaulting to certain motifs. Style incongruencies occurred when text or elements otherwise incongruent with the style would emerge in the middle of the generation. The area shaded in green shows reference images of styles that give context for why certain generations were failures. }
	\Description{Pictured in this image is multiple rows delineating all the failure modes expressed in the results of Experiment 4 for the style examination. These rows include "meaning misunderstood", "inability to capture style", "incongruency within style" from emergent photos, "incongruency within style" from emerging text, and defaulting to stereotypical motifs. There is a green shaded region that shows reference images of what a successful generation might have included. For example, in the first row, there are examples of the word mola being misinterpreted not as an art style but of an animal.}
	\label{fig:bad}
\end{figure*}
\subsubsection{Failure mode: Style misunderstandings due to the multiplicity of meanings in text.}

As mentioned in the previous section, the model interpreted Baroque and Neoclassical styles through the lens of decor, architecture, and sculpture, generating motifs from Baroque furniture and Neoclassical architecture and sculpture as opposed to Baroque and Neoclassical painting.

Many of the styles that performed the most poorly from the annotations were misunderstood by VQGAN+CLIP in some dimension. For example, dark academia, a social media aesthetic captured by a romanticized, Gothic approach to esoteric motifs often returned generations that contained components of a cartoon character under dramatic lighting. One possible explanation is that the model was influenced by another popular entity on the Internet that also involved the word 'academia'--the anime \textit{My Hero Academia} (the characters emerging in many of those generations shared a distinct green hair color).

Another case was the style of mola, a Latin American folk art form from Latin America with a vividly saturated color palette and heavily stylized characterization of subjects. Mola the art style was misinterpreted as mola, a species of fish. All generations of mola had a predominantly blue color palette that evoked something aquatic, and many of the subjects were blended to look like fish ("see man in the mola style" in \autoref{fig:bad}). A potential cause for understanding mola as a fish species could be attributed to the bias towards animal species from ImageNet1000, a significant subset of which were animal species. However, the mola that were represented also were not photorealistic but rendered in a stylized form. 
These examples represent how conflicting interpretations of a prompt can lead to misinterpretation within the generated image. 
Misunderstandings could also arise from different \textit{parsings} of the prompt. For example, take the action painting art style, which is meant to refer to artists who painted dynamically using random drips and splatters. When the model generated for "a man" or "woman" in the action painting style, it created a generation that implied a \textit{man in action} or a \textit{woman in action}.

The many different instantiations of this failure mode suggests that the multiplicity of meanings within language both at the word level and the sentence level can present as a problem for text-to-image generation. It also represents a fundamental shift in the thought process behind the creation of a visual artwork. Visual artwork usually involves thinking about the spatial specifics of the composition, which text-to-image generation does not lend well to.

\subsubsection{Failure mode. Inability to capture styles in a complete sense}

Another theme within unsuccessful styles was the inability to express styles that were more symbolic than visual. For example, Dadaism, a style representing the rejection of capitalism and embrace of the avant-garde and nonsense, was rated very poorly with a mean subjective rating of 1.42. Dadaism traditionally was expressed through satire and collaging, and it tended to involve cultural knowledge and nuanced symbolism relating to pop culture and politics.
Likewise, Bauhaus was another abstract style that was heavily influenced by abstract values like harmony and utility rather than visual abstraction. While Bauhaus has a characteristic visual style rooted in geometric shapes, much of that is illustrated in architecture as opposed to image.
These styles and their poor performance in the annotation study illustrate there are still abstractions and pools of cultural knowledge that are either not well understood or visually representable within text-to-image-models, potentially because of their different angle of abstractness. (An example of "happiness in the Dadaism style" is shown in row 4, column 1 of \autoref{fig:bad}.)

Other styles were simply insufficiently captured. For example, if we refer to any of the images of Kerala mural style generations in \autoref{fig:bad}, we see that they never reached the vividly saturated and stylized look of actual Kerala mural frescos. However, they approached it, evoking color combinations and motifs such as traditional dress that established an association of Kerala murals with India.

\subsubsection{Failure mode. Style incongruency, often in the form of emerging photos or text}

Sometimes elements that interrupted the style would come through the generation. These elements could be bucketed into two cases: photorealistic elements or text elements.

For example, in \autoref{fig:bad} we can see a cat done in the style of Cubism. The cat's fur is entirely photorealistic, which is out of place in an otherwise abstract image. Likewise, in images of river, a photorealistic texture of a river surface would often surface even if the style was intended to be a sketch. This phenomena tended to occur for concrete subjects such as dogs, cats, rivers, and oceans.

The second case was when text began to emerge within images across iterations. For example in the generation "happiness in the Dadaism style", we see "DADA" explicitly written out across the generation, potentially as a compensatory technique on the model's part to optimize towards the prompt. Curiously, for ASCII art, text never manifested, and each generation was composed of similarly sized blurs of alphanumerical literals.

\subsubsection{Failure mode: Defaulting to motifs.}

Another failure mode expressed within styles such as unreal engine and High Renaissance was a defaulting to certain motifs. For example, for the style unreal engine, most of the text prompts returned a scene with textured rocks and grass akin to what might be rendered by a game engine.

\subsection{Results and Discussion of Partitions}

\subsubsection{Abstract versus figurative}

\begin{figure*}
  \centering
  \includegraphics[width=\linewidth]{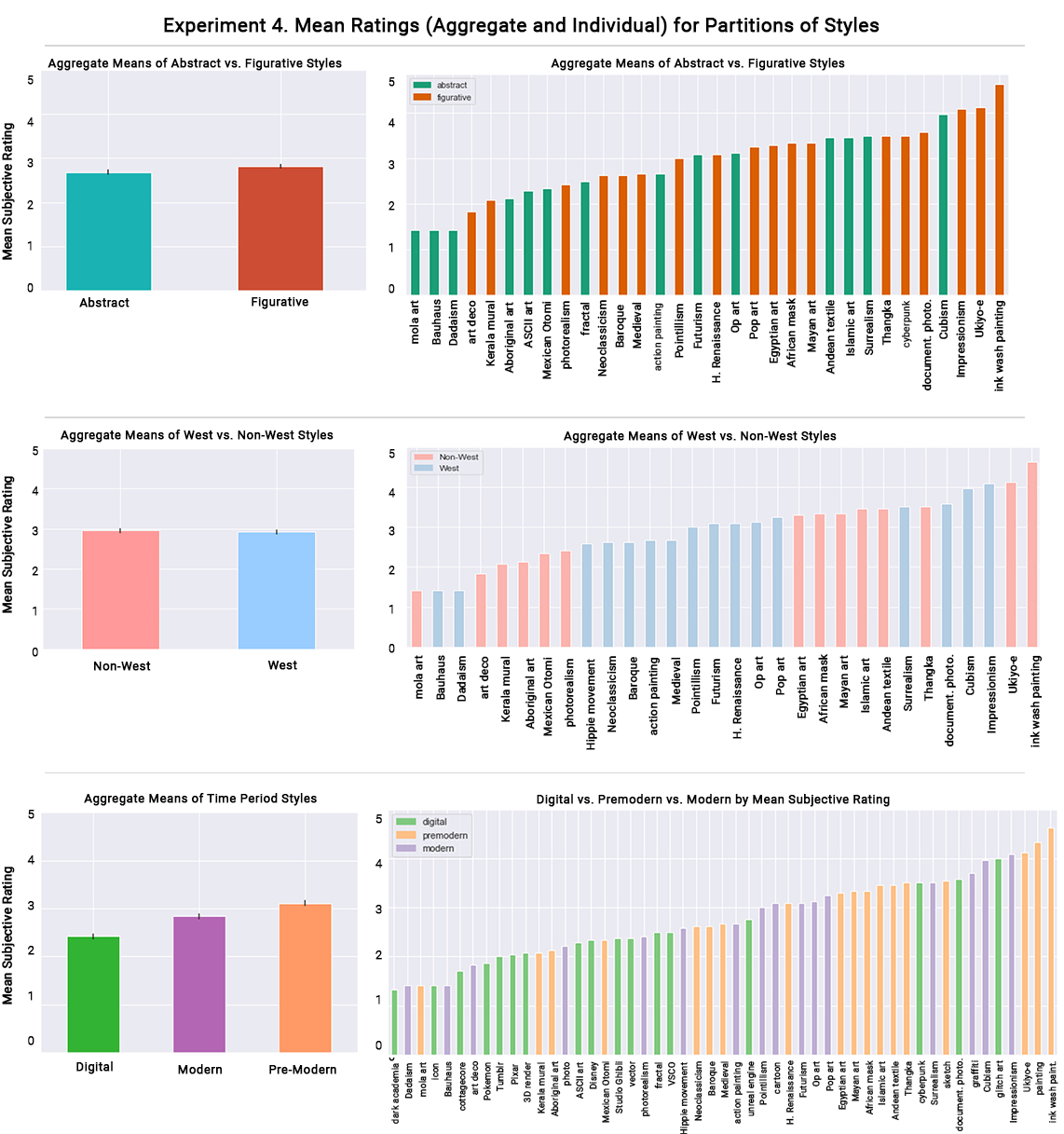}
  \caption{For Experiment 6, in the left subgraphs, averages are reported for each category of the three partitions studied: (Abstract, Figurative), (West, Non-West), (Digital, Modern, Premodern). The right figures are bar graphs which rank each style included in the partitions by their aggregate means from low to high left to right, coloring for their respective categories. We found significant differences between the abstract and figurative styles in aggregate as well as the digital, modern, and pre-modern styles in aggregate.}
  \Description{Pictured are aggregate graphs for the different partitions studied in Experiment 4: abstract-figurative, non-western versus western, and time periods digital, modern, and premodern. There are bar graphs for each partition showing the performance of each style colorcoded for their categorization along the partition. }
  \label{fig:stylepartition}
\label{fig:fig}
\end{figure*}

\begin{figure*}[t]
	\centering
\includegraphics[width=\linewidth]{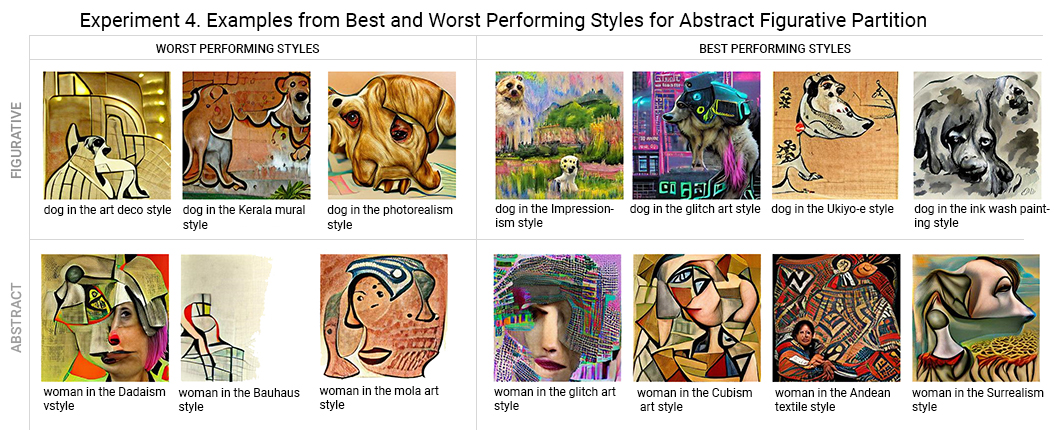}
	\caption{For Experiment 4, we illustrate some of the best and worst styles along the abstract and figurative style partition.}
	\Description{Pictured are some example generations of the best and worst styles along the abstract and figurative style partition.}
	\label{fig:absfig}
\end{figure*}

To investigate whether the model performed better on abstract styles or figurative ones, we looked at a subset of 33 specific styles, excluding styles such as more general mediums (i.e. a painting, a photo) and Internet aesthetics (i.e. dark academia). We looked at a subset because certain styles generally did not fall cleanly between abstract and figurative styles.

We found that abstract styles averaged a 2.63 rating (standard error 0.06) , while figurative styles averaged a 3.16 rating (standard error 0.06) . After running a chi squared test on the frequencies of ratings we found the difference between these ratings was significant to a p value of < 0.01. In \autoref{fig:absfig}, we visualize the top 4 styles and worst 3 styles for abstract and figurative styles. In \autoref{fig:stylepartition}, we color code the ranked styles by their abstract or figurative nature.

Our original hypothesis was that abstract styles would perform better because we thought they would be more tolerant to the deconstructed, global incoherence of many generations. We found that while our original hypothesis was correct for styles such as glitch, Cubism, and Andean textiles for reasons we expected (such as a high tolerance for deconstruction), abstract styles were prone to a wide range of failure modes. These failure modes included misunderstandings due to misinterpretation and an inability to access higher-order cultural knowledge.

The figurative styles that performed in the top 4---Ukiyo-e, Impressionism, documentary photography, and cyberpunk---displayed a diverse range of stylistic details from line to texture to perspective. The worst performing figurative styles suffered from different modes of failure such as an inability to capture the style (Kerala mural style generations) or a defaulting to unconvincing motifs (as in the case of the generation \textit{dog in the art deco style} seen in \autoref{fig:absfig}).

\subsubsection{Western versus non-Western}

\begin{figure*}[t]
	\centering
	\includegraphics[width=\linewidth]{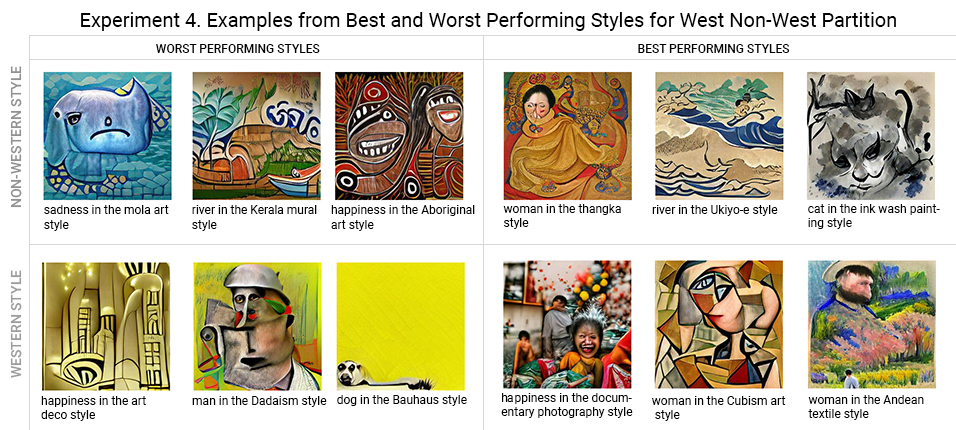}
	\caption{For Experiment 4, we illustrate some of the best and worst styles along the Western and Non-Western style partition.}
	\caption{Pictured are some of the best and worst styles along the Western and Non-Western style partition.}
	\label{fig:westnonwest}
\end{figure*}

To investigate whether the model would perform better on Western or Non-Western art styles, we looked again at a subset of specific styles, excluding mediums and Internet aesthetics (as they tended to be more globalized).

We found that Western art styles averaged 2.92 (standard error: 0.07), while non-Western art styles averaged 2.95 (standard error: 0.06). Using a Mann-Whitney test, we found that there was an insignificant difference between the distribution of ratings for Western styles and the distribution for ratings for non-Western styles (p-value: 0.377).

We illustrate the top performing styles in \autoref{fig:stylepartition}, where we show the ranking colored for Western for Non-Western styles. Our findings suggest that the difference between Western and Non-Western styles was actually insignificant. One straightforward reason is that the Internet scale data could have compensated for the relative obscurity of any style. The alternating and even spread of the Western and non-Western styles over the x-axis of the bar graph illustrating individual styles by ranking is also visually suggestive of the same result.

\subsubsection{Time period: premodern, modern, versus digital}

\begin{figure*}
	\centering
	\includegraphics[width=\linewidth]{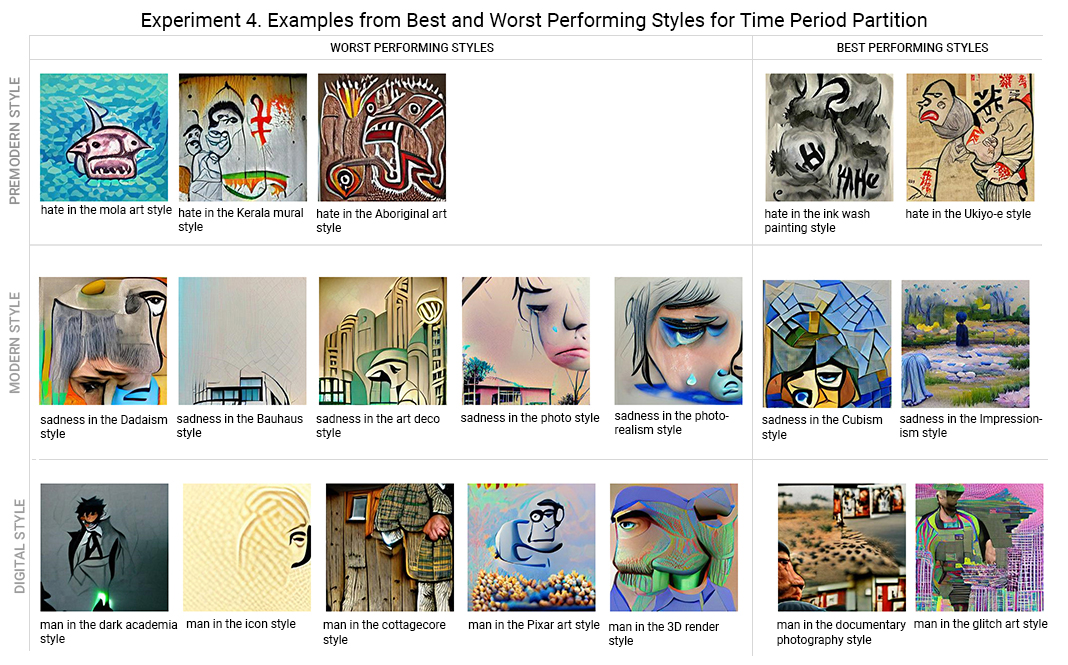}
	\caption{For Experiment 4, we illustrate some of the best and worst styles along the style partition for different time periods: digital, modern, and premodern.}
	\caption{Pictured are some of the best and worst styles along the time periods partition: digital, modern, and premodern.}
	\label{fig:digital}
\end{figure*}
We investigated whether the model would perform better on digital styles (Internet aesthetics), modern, or premodern art styles. We partitioned the styles into these time periods and colored these ranked styles by category in \autoref{fig:stylepartition}.

We found that digital styles performed the worst, then modern styles, and then premodern styles with aggregate annotator ratings of 2.41, 2.83, and 3.11 respectively. Using a Kruskal Wallace test, we found these differences to be highly significant p-value < 0.001. One potential reason for why digital styles performed the worst could be that the digital styles we covered had more inherent stylistic range. Some digital styles such as Tumblr, could be represented by multiple photo filter palettes, while others such as \textit{cottagecore} and \textit{dark academia} could be represented through different aesthetic forms (i.e. an outfit in fashion, a piece of furniture). Still other styles like Disney encompassed a range of visual styles within itself, even though the generations came across colors and lines reminiscent of the Disney Renaissance.

Given the results from \autoref{fig:allstyles}, we can see that the model is able to capture an extensive range of styles even if it performs differently dependent upon the nature of the style. Many perform well so long as they are not prone to misinterpretation or other aforementioned failure modes. We conclude from this experiment the following design guideline: \textbf{when choosing the style of the generation, feel free to try any styles, no matter how niche or broad.} 

\section{Experiment 5: Interaction between subject and style}

Given the varied but still successful application of style as a steering keyword within prompts, we wanted to investigate the subject keywords similarly and then observe how subject and style as parameters would interact with each other. We first ran a pilot experiment studying subject alone. However, we chose not to take this experiment further, because the generations yielded were too consistently poor due to the underconstrained nature of the prompt. See the \autoref{fig:subjectonly} in the Appendix for further examples of this pilot.
We focus on the interaction of subject and style in this experiment, and pursue the following research questions: \textbf{To what degree do categories of subject and style influence one another? Do categories of styles, such as abstract or figurative styles, perform better on certain categories of subjects, such as abstract or concrete subjects?}

\subsection{Methodology}

To study the effect of interaction of subject and style, we generated 1581 images from 51 subjects, 31 styles.  The full list of subjects and styles are in the Appendix, but follow the same rationale as previous experiments for coverage across the abstract-concreteness spectrum (for subjects) and diversity of styles in terms of time, schools of art, and levels of abstraction.

\subsection{Annotator Methodology}
 We recruited two annotators who had domain knowledge in art and design respectively for this task. Each received a set of subject and style combinations, ordered in different random order. They annotated each generated image for the coherency of subject and style within the image as per the following rubric:

\begin{itemize}
	\item 1: Extremely poor representation of subject and style
	\item 2: Bad representation of subject and style
	\item 3: Average representation of subject and style
	\item 4: Good representation of subject and style
	\item 5: Excellent representation of subject and style
\end{itemize}

\begin{figure*}[t]
  \centering
  \includegraphics[width=0.9\linewidth]{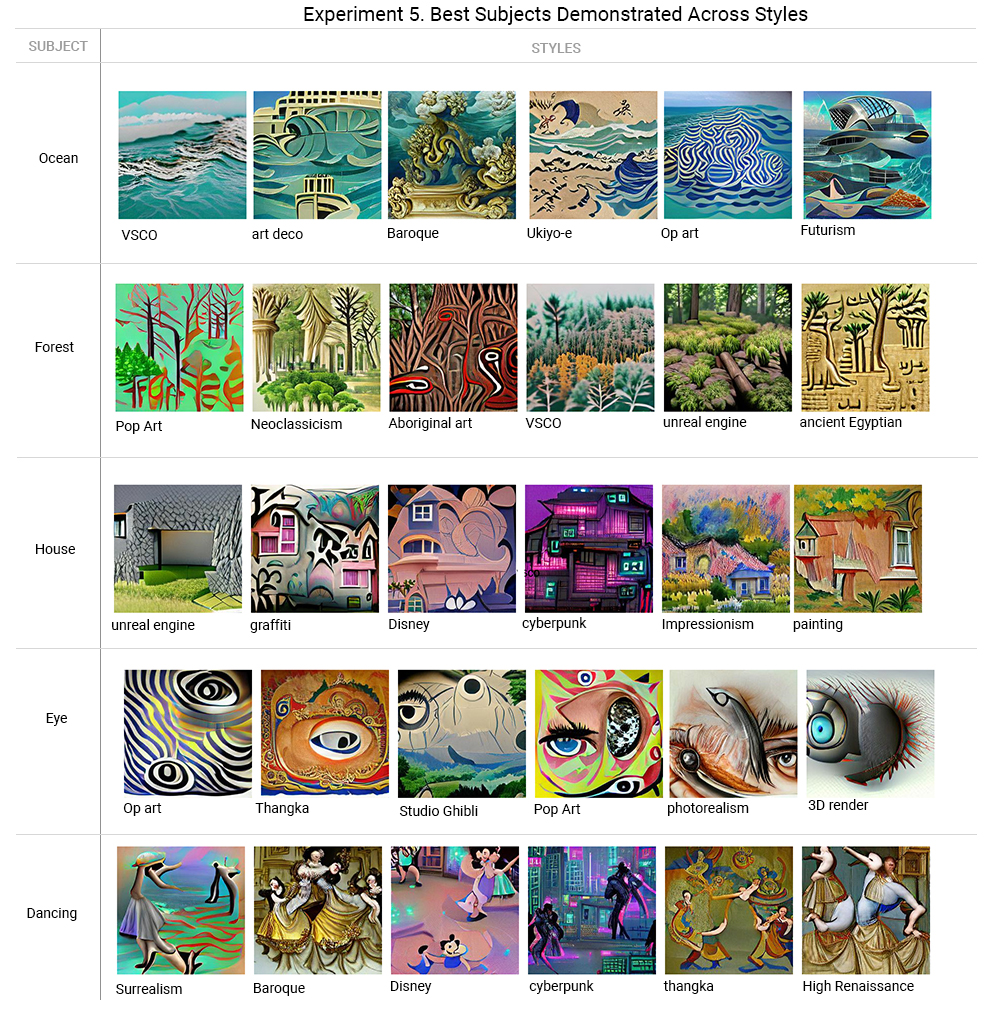}
  \caption{For Experiment 5, generations from the top five subjects of 51 subjects are visualized above in various styles.}
  \Description{Pictured are rows of generations for the top five subjects of Experiment 5. These subjects were: ocean, forest, house, eye, and dancing. The rows are 6 generations long, comprising various styles.}
  \label{fig:topsubjectstyle}
\end{figure*}

\begin{figure*}
  \centering
	\includegraphics[width=\linewidth]{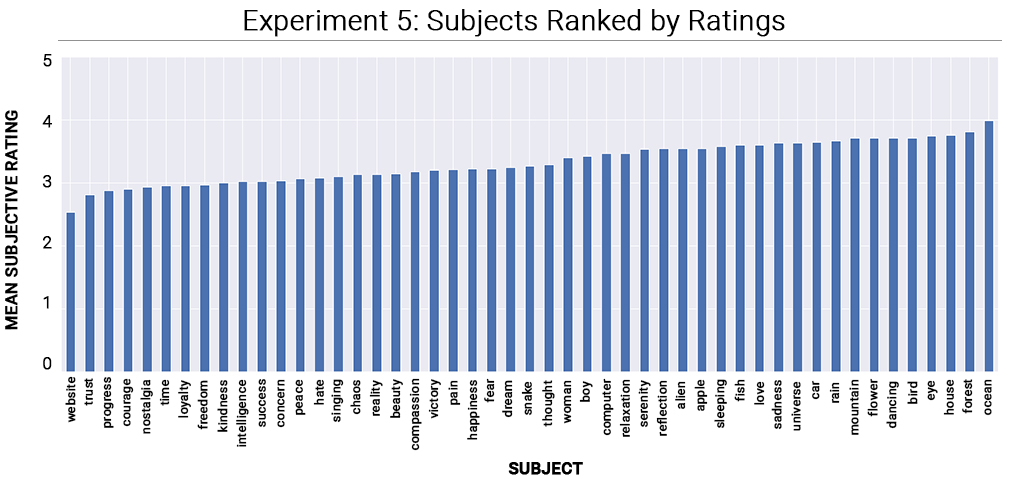}
  \caption{For Experiment 5, 51 subjects were crossed with 31 styles. When mean rankings were aggregated across styles, the top 10 subjects all were concrete subjects. The top five specifically were ocean, forest, house, eye, and dancing.}
  \Description{A bar graph illustrates how subjects of various degrees of abstractness fared in an annotation task for Experiment 5. The X axis is 51 subjects, and the Y axis is the mean subjective rating across annotators.}
  \label{fig:subjectaggregate}
\end{figure*}

\subsection{Results}

Two variables we wanted to test in the experiment were the abstract or concrete nature of the subject and the abstract or figurative nature of the style.

We first studied just the abstract or concrete nature of the noun alone, aggregating results by subject. We found that the top ten subjects were all categorically concrete, with an average concreteness value of 4.47. They were all subjects that were universal across most cultures: ocean, forest, house, eye, bird. Examples of these top subjects crossed with different styles are illustrated in \autoref{fig:topsubjectstyle}. We found that when we compare the abstractness of the noun to the quality of the generation, there is an r value / Pearson's coefficient of 0.62, which implies a moderate to strong positive association. This means that on average there is a trend where concrete subjects tend to do better.

We then considered the influence of the abstract or figurative nature of style as well, by looking at the generations from a factorial 2x2 lens. We found the following aggregate rankings for the enumerated categories: abstract-abstract (3.05), abstract-concrete (3.17), figurative-abstract (3.49), and figurative-concrete (3.54). In running a two-way ANOVA on the annotations we found that all p-values were significant, being well below 0.01. This allows us to conclude that both factors have a significant effect on the rating of the generation. Likewise, we saw that their interaction is also significant to a p-value well below 0.01.

\subsection{Success and Failure Modes for Subjects Crossed With Styles}
In the following section, we perform a qualitative analysis on what success and failure modes we observed for subject-style generations.

\begin{figure*}[t]
	\centering
	\includegraphics[width=\linewidth]{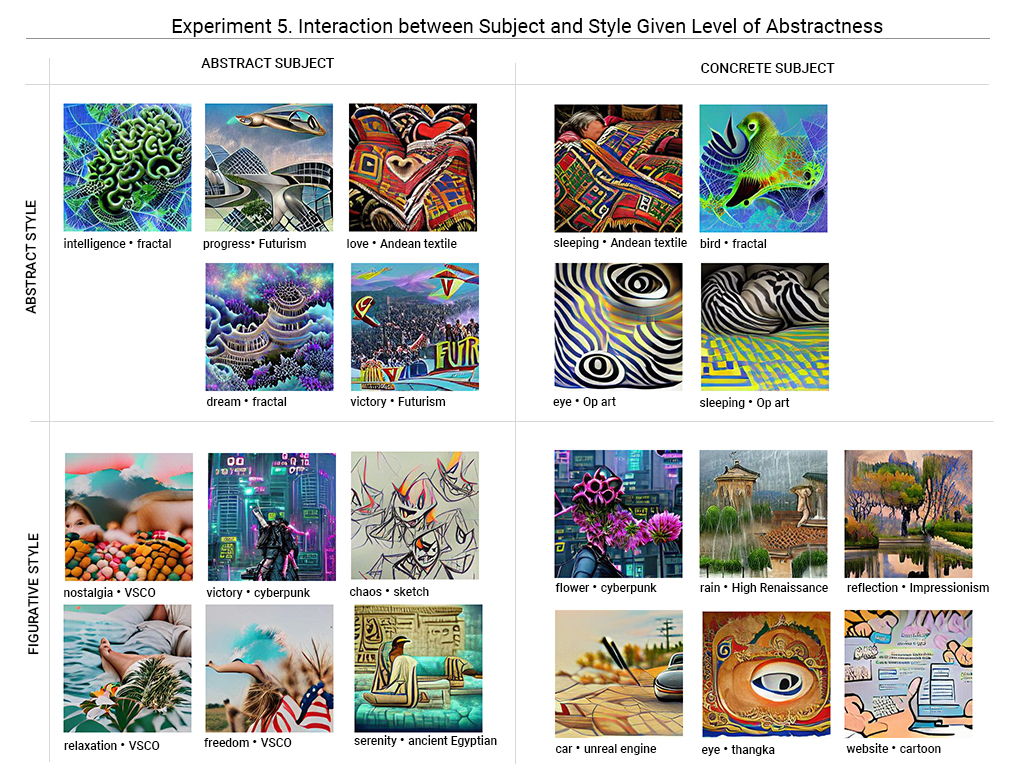}
	\caption{In Experiment 5, we looked at both subject and style words in the prompt. We crossed abstract and concrete subjects with abstract and figurative styles. In this figure above, we show some success cases within each crossed category. In running a two-way ANOVA, we found that both subject and style have a significant effect on the rating of the generation. Likewise, their interaction was also statistically significant.}
	\Description{Pictured is a quadrant plot of different generations. In the top left is a quadrant for abstract style and abstract subject generations. In the top right is a quadrant for abstract style and concrete subject generations. In the bottom left is a quadrant for figurative style and abstract subject generations. In the bottom right is a quadrant for figurative style and concrete subject generations. These generations are ones that successful capture both subject and style in the image.}
	\label{fig:substyle}
\end{figure*}

\subsubsection{Success mode. Correct applications of symbolism}

In many subjects, the text-to-image framework was able to demonstrate that it could access and apply symbols. For example, in most generations for hearts, heart symbols emerged out of the image (even if the symbol was incongruent with the style, for example as a heart symbol would be in Ukiyo-e art).

However, generations also showed a flexible understanding of love in the form of kisses, proposals, and hugs. Generations in the subject of sadness also demonstrated an expressive range of symbols for sadness such as blueness, frowns, tears, and lonely figures. For other abstract subjects such as freedom, relaxation, or serenity, the model was able to demonstrate that it could connect freedom with American flags and relaxation with reclining. These associations are intuitive, even if certain connections, such as freedom with the United States, have overtones of bias.

This success mode is primarily what makes the difference between good generations and bad generations for abstract subjects. A generation from an abstract subject is successful only when it is able to find purchase in the image as a symbol. Using a symbol to stand in for an abstract subject is apparent in both abstract and figurative styles.

\subsubsection{Success mode. Integration of motifs with elements of the subject }

Another mode of success that we could see in generations from \autoref{fig:topsubjectstyle} such as "eye in the style of Op art" from the abstract style, concrete subject category or "intelligence in the fractal style" from the abstract style, abstract subject category was when components of the subject and style matched and blended well. In the "intelligence in the fractal style", intelligence is symbolized in a brain which has recursive convolutions of gray matter, which elicits the idea of a brain is a fractal.
Other examples such as "flower in the cyberpunk style" or "nostalgia in the VSCO style" in the other quadrants of \autoref{fig:substyle} demonstrate how the color palette of a style colored the subject. In the former, the flower took on the magenta trademarks of cyberpunk and in the later, nostalgia was established through a sepia and pastel tones reminiscent of filters.

What makes generations with concrete subjects successful is when the subject is able to emerge from a style without disrupting it. For example, in the generation \textit{rain in the High Renaissance} style in \autoref{fig:substyle}, we see that the rain is pervasive but drawn in fine, white strokes that are characteristic to the style. Likewise, we see that the \textit{car in the unreal engine style} image applies the same effects of depth of field and scene lighting prevalent in all CG renders.

\begin{figure*}[t]
	\centering
	\includegraphics[width=\linewidth]{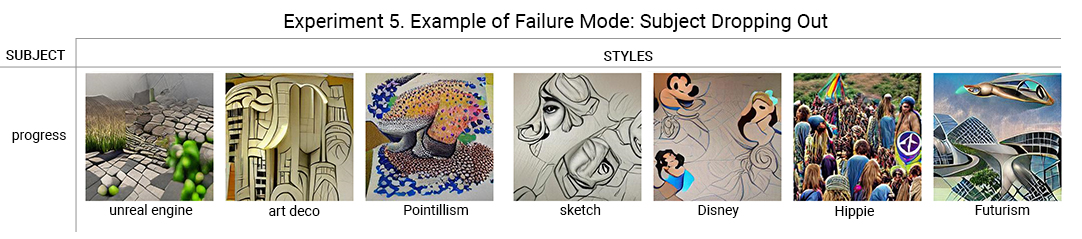}
	\caption{Experiment 5. We use the subject "progress" here to illustrate the failure mode where the subject drops out for certain styles (see left three images). However, nuances of progress were conveyed nonetheless in the right four images if we consider progress through different definitions such as progress pictures, progressivism, and something evocative of the future.}
	\Description{Pictured are generations of the subject progress with various styles. Four on the left in the row are poor examples, where the subject drops out. Three on the right are examples where different nuances of the word progress are captured in the image. }
	\label{fig:progress}
\end{figure*}

\begin{figure*}[ht]
	\centering
   \includegraphics[width=\linewidth]{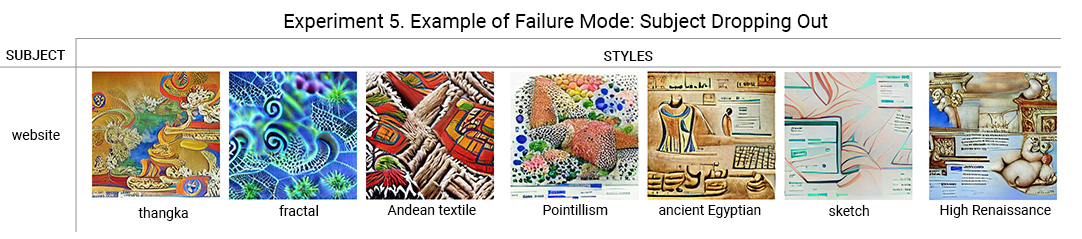}
	\caption{Experiment 5. We use the subject "website" to exhibit when the subject would drop out with certain styles. This subject represented a challenge because it was an anachronistic subject for most styles. }
	\Description{Pictured are generations of the subject website with various styles.  }
	\label{fig:website}
\end{figure*}

\subsubsection{Failure mode. Subject dropping out}

A common failure mode, particularly for abstract subjects, was when the subject would not come through. For example, one of the most poorly rated subjects was \textit{progress}. This is understandable, because progress is a difficult word to visualize. The images on the left side of the row illustrate little relevance to the subject of progress. However, it does not mean that progress was not picked up at all by the model. The right four generations show potential interpretations of the word progress relevant to the styles they were generated in: Disney, Futurism, and Hippie; progress was captured in the respective styles as progress pictures, a vision of the future, and potentially the progressivism and activism associated with crowds. Therefore even subjects that the model does poorly on with some styles can show nuanced understanding if there is some relevance suggested by the combination of subject and style.

The most poorly rated of the subjects was \textit{website}. This one is interesting because it represents a challenge to the framework, because websites and digital media are anachronistic subjects to many modern and all premodern styles. We found that the model sometimes simply dropped \textit{website} from the generated image, which we would say is neither correct nor incorrect as the subject could have conflicted with the style. This is another outcome that suggests that relevancy should be a consideration for users interacting with text-to-image generative frameworks. However, in these generations we could also see positive outliers where the style adapted to the subject. For example, for "website in the ancient Egyptian style" seen in \autoref{fig:website}, we can see a keyboard, and a person using a computer screen.

\subsubsection{Failure mode. Nightmare fuel and mature images}
Another failure mode was when the model returned images that were either grotesque or inflammatory in nature. For example, in images using the style of High Renaissance, a common motif would be dramatic renderings of human bodies in shadow. However, perhaps because the model is in part convolutional, the motif of muscled body parts was seamlessly repeated within images until the image was cluttered with bulbous, grotesque, and interconnected limbs.

Likewise, when the authors ran Internet aesthetics as styles during pilot experiments and put in prompts such as "girl in the style of Tumblr" or "hate in the style of VSCO", the model returned images that were reminiscent of pornography and self harm. The prompts innocuously requested images for which the model added specific, unsettling details without giving any system feedback in the form of trigger warnings.

The repetitive motifs suggest that trigger warnings should be put into place for pareidolia (our subconscious tendency to read emergent forms from parts and patterns) and maturity warnings. Text-to-image generation likely translates the biases learned from the Internet into imagery. These images are excluded from figures for the sake of propriety.

In summary, for Experiment 5, we concluded the following design guideline: \textbf{When picking the subject of the generation, pick subjects can complement the chosen style in level of abstractness and relevance.}

\section{Discussion}
In a series of experiments, we demonstrated that a range of "SUBJECT in the style of STYLE" generations can be arrived at quickly and easily with a text-to-image generative framework. We looked at different parameters for prompt engineering such as subject and style (Experiments 1, 4, 5) and studied the effects of modulating hyperparameters like the number of iterations and random initializations (Experiments 2, 3). 

We condense our findings from the previous experiments into design guidelines and results to elaborate default parameters and methods for end users interacting with text-to-image models.

\begin{itemize}
	\item \textbf{When picking the prompt, focus on subject and style keywords instead of connecting words.} Rephrasings using the same keywords do not make a significant difference on the quality of the generation as no prompt permutation consistently succeeds over the rest.
	\item \textbf{When generating, generate between 3 to 9 different seeds to get a representative idea of what a prompt can return.} Generations may be significantly different owing to the stochastic nature of hyperparameters such as random seeds and initializations. Returning multiple results acknowledges this stochastic nature to users.
    
	\item \textbf{When generating, for fast iteration, using shorter lengths of optimization between 100 and 500 iteration is sufficient.} We found that the number of iterations and length of optimization did not significantly correlate with user satisfaction of the generation.

	\item \textbf{When choosing the style of the generation, feel free to try any style, no matter how niche or broad.} The deep learning frameworks capture an impressive breadth of style information, and can be surprisingly good even for niche styles. However, avoid style keywords that may be prone to misinterpretation.

	\item \textbf{When picking the subject of the generation, pick subjects that can complement the chosen style in level of abstractness.} This could be done by picking subjects for styles considering how abstract or concrete both are or pairing subjects that are easily interpretable or highly relevant to the style.
	\item \textbf{When looking at the results, present users with trigger warnings for pareidolia and offensive content.} The models currently do not acknowledge the possibility for offensive content.

\end{itemize}
Our experiments gave us empirical grounding to focus many of the hyperparameters and free parameters for prompts that otherwise make prompt engineering and text-to-image generation otherwise overwhelming, unbounded, and inexhaustive.

\subsection{Implications of Borrowing Styles}

While text-to-image interaction presents a novel and emerging form of human-computer interaction for media creation, this advancement presents us with new sets of concerns. Suggesting that we use pre-existing styles is at once intuitive and controversial. There are many implications to borrowing styles as keywords, one of which is that we are relying on a machine's non-expert understanding of a style to generate outputs. This makes it possible for text-to-image models to return generations that could err towards stereotypes and other misrepresentations.

For example, one of the top three styles in terms of ratings for Experiment 4 was Ukiyo-e. These generations tended to employ beige, black, and muted primary colors suggestive of woodblock prints. However, Ukiyo-e work in the past was not confined to this range of color. Ukiyo-e as a style spanned centuries, during which as a style it exhibited different approaches to color ranging from monochromatic ink to brilliant brocades. This implies that the model could only shallowly summarize Ukiyo-e. Likewise, sketches tended to return black and white images, belying an stereotypical understanding of sketches as generally black and white.

While styles can be borrowed as keywords to prompt generations, styles can also be misrepresented. For example, in our analysis of failure modes for styles in Experiment 4, we found that certain styles could easily be misinterpreted by the framework for alternative homonym meanings. For example, Neoclassical generations often had motifs from Neoclassical architecture but not Neoclassical painting. It could also be argued that the generations of many styles only flatly reproduced the styles (with \textit{approximate} colors and technique), and that generations have the potential to add unwelcome noise to the bodies of work behind these artistic traditions.

Styles could also refer to specific individual names (i.e. a painting in the style of Picasso) , and while this is something we did not formally explore in the paper, it is a method people have used in the wild (i.e. prompting with a subject in the style of a specific artist). Generations in these categories warrant discussion about copyright and appropriation of existing material.

\subsection{Limitations and Future Work}
Our focus in this paper was on prompt engineering a text-to-image framework with text. However, the model could have also received images to start optimizing from. We believe studying how the model can be conditioned on an image and text together is interesting future work that could provide insights into how people can move between different modes of interaction. Interacting with text is high-level interaction while working with images is low-level and more conducive to directly manipulating the generation. Similarly along the lines of user control, another line of work would be to improve the capacity of this framework for iteration. Currently, users can only regenerate upon rerunning the framework on previous generations, but usability could be improved if more controls for steering the generation at intermediate stages could be exposed \cite{10.1145/3313831.3376739}.

Our qualitative analysis in both Experiment 4 and 5 demonstrates that more work could be done to explore the nuances of what certain styles can elicit. Styles exist with respect to cultural contexts and histories, and it is valuable to understand how generations can be pushed to be more than flat reproductions of styles. For example, one could say that generations in the style of Impressionism or Cubism could emulate these respective styles at least at the surface level, in terms of technique and color palettes. However, it remains to be explored to what degree could these generations channel the nuances of these styles, such as their conceptual values or messages.

Another limitation of this paper is that for most of the experiments, we only looked at one prompt ("SUBJECT in the STYLE of") for VQGAN+CLIP. We looked at this prompt and framework because it had traction within creative technologist communities, but further research could look into other prompts and models. For example, what would happen if we typed in the first line of a poem, a news headline, or a design goal for a moodboard? Additionally, for this prompt and others, there are modifiers that we could have explored to increase the realistic quality of the generation. For example, we could have added and systematically explored modifiers like "4k" or "2048px". 

Given that text-to-image generation is an emerging paradigm of interaction, there are many avenues of prompt engineering for visual generation tasks that future work can explore. 

\section{Conclusion}
In this paper, we conducted a series of five experiments that each tackled a different angle of prompt engineering for text-to-image generative models involving prompt permutations, random seeds, length of optimization, style keywords, and subject and style keywords. Our experiments found significant differences between the quality of generations that fell into different categories of style as well as subject and style. We summarized the failure and success modes of these generations through qualitative close reads of these generations. Additionally, we empirically found ranges for hyperparameters where their effect was significant, such as in the case of seeded random initialization and length of optimization. From these experiments we were able to synthesize design guidelines to guide users through the unbounded, stochastic, and prone-to-error nature of text-to-image interaction.

\begin{acks}
Vivian Liu is supported by the NSF Graduate Research Fellowship (DGE-1644869).
\end{acks}

 \bibliographystyle{ACM-Reference-Format}
\bibliography{citations}

\newpage

\appendix
\section{Experiment 1. Prompt Permutations}

We chose the listed prompt permutations for the following reasons:
\begin{itemize}
   \item \textbf{A MEDIUM of SUBJECT in the STYLE style }  --- We wanted to test this prompt because the authors of CLIP noted that incorporating MEDIUM words could help return better generations \cite{openai}. For example, inputting a prompt such as a "a painting of a dog in the Cubism style"  would lead to better results than "dog in the Cubism style".
\item \textbf{A STYLE MEDIUM of a SUBJECT} --- We wanted to test this prompt because it was a reordering of prompt permutation \# 1.
\item \textbf{SUBJECT STYLE} --- We wanted to test this prompt because it was the most minimal amount of information to prompt the machine with. At the same time, this sort of text prompt is how we regularly query image search engines.
\item \textbf{SUBJECT. STYLE} --- We wanted to test this prompt, a close cousin of prompt permutation \#3 and observe the effect of punctuation.
\item \textbf{SUBJECT in the style of STYLE} --- We wanted to test this prompt because it had traction within the creative technologists community \cite{komatsuzaki_2021}.
\item \textbf{SUBJECT in the STYLE style} --- We wanted to test this prompt because it was a rephrasing of prompt permutation \#5.
\item \textbf{SUBJECT VERB in the STYLE} style --- We wanted to test this prompt permutation to see the influence of verbs. The authors of CLIP in their repository noted that the model performed better on nouns than verbs given the noun-centric supervision of Imagenet.
\item \textbf{SUBJECT made/done/verb in the STYLE art style} --- Example: We tested this prompt permutation as a rephrasing of prompt permutation \#7.
\item \textbf{SUBJECT with a STYLE style.} --- We tested this prompt permutation to test a different ordering with a different function word.

 \end{itemize}

\section{Experiment 3. Random Seeds}

We chose to expand our list of styles to include a broader range of concreteness values, such that they included: \textit{love, hate, peace, progress, relaxation, loyalty, compassion, beauty, pain, dream, thought, trust, freedom, chaos, success, courage, happiness, nostalgia, intelligence, kindness, time, concern, sadness, reality, serenity, fear, victory, happiness, alien, car, house, apple, singing, dancing, sleeping, mountain, rain, ocean, forest, flower fish, bird, snake, boy, woman, eye, computer, website, universe, reflection}. Likewise, we chose to use the same set of styles: \textit{photorealism, Studio Ghibli, Neoclassicism, African mask, thangka, fractal, Hippie movement, ancient Egyptian, art deco, unreal engine, Disney, cartoon, Pop art, VSCO, Futurism, 3D render, Pointillism, sketch, Surrealism, Andean textile, Aboriginal art, Ukiyo-e, High Renaissance, Mayan, graffiti, Cubism, Impressionism, Baroque, Op art, cyberpunk, and painting}.
\section{Experiment 5. Subjects}

\begin{figure*}[t]
	\centering
	\includegraphics[width=\linewidth]{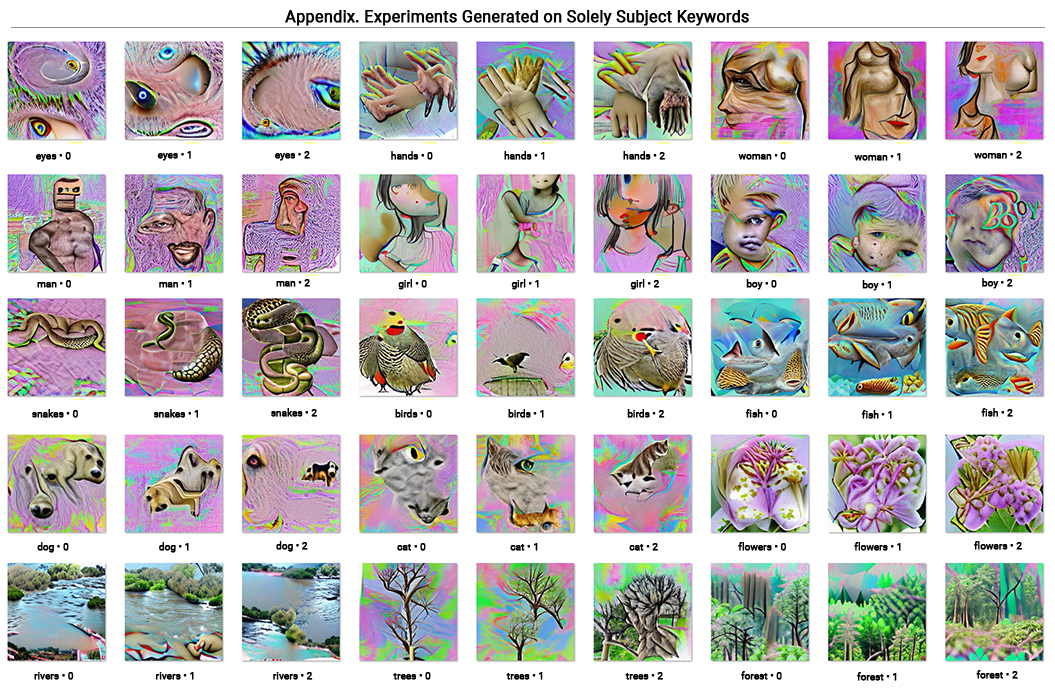}
	\caption{In a pilot before Experiment 5, we found that using only subjects as keyword dimensions was insufficient. The underconstrained nature of the generation made generations too poor to evaluate, because they were not grounded in any sort of aesthetic.}
	\Description{Pictured is a grid of subjects generated with three different seeds. Each one looks deconstructed and the overall aesthetic is a noisy color palette.}
	\label{fig:subjectonly}
\end{figure*}

\end{document}